\documentclass[11pt,
twocolumn,
 aip,
 apl,%
 showkeys,
 amsmath,amssymb,
reprint
]{revtex4-1}
\usepackage[breaklinks=true,colorlinks=true,linkcolor=blue,urlcolor=blue,citecolor=blue]{hyperref}
\usepackage{graphicx}%
\usepackage{graphics}%
\usepackage{epstopdf}
\usepackage{dcolumn}%
\usepackage{soul,color}
\usepackage{bm}
\usepackage[mathlines]{lineno}
\expandafter\ifx\csname package@font\endcsname\relax\else
 \expandafter\expandafter
 \expandafter\usepackage
 \expandafter\expandafter
 \expandafter{\csname package@font\endcsname}%
\fi
\usepackage[utf8]{inputenc}
\usepackage[T1]{fontenc}
\usepackage{mathptmx}

\usepackage{siunitx} 
\DeclareSIUnit{\belmilliwatt}{Bm}
\DeclareSIUnit{\dBm}{\deci\belmilliwatt}

\usepackage{miller} 

\begin{document}

\preprint{AIP/123-QED}

\title[Long-range spin-wave propagation in transversely magnetized nano-scaled conduits]{Long-range spin-wave propagation in transversely magnetized nano-scaled conduits}

\author{Björn Heinz}
\email{bheinz@rhrk.uni-kl.de}
\affiliation{Fachbereich Physik and Landesforschungszentrum OPTIMAS, Technische Universität Kaiserslautern, D-67663 Kaiserslautern, Germany}

\author{Qi Wang}%
\affiliation{Faculty of Physics, University of Vienna, A-1090 Wien, Austria}

\author{Michael Schneider}%
\affiliation{Fachbereich Physik and Landesforschungszentrum OPTIMAS, Technische Universität Kaiserslautern, D-67663 Kaiserslautern, Germany}

\author{Elisabeth Weiß}%
\affiliation{Faculty of Physics, University of Vienna, A-1090 Wien, Austria}

\author{Akira Lentfert}%
\affiliation{Fachbereich Physik and Landesforschungszentrum OPTIMAS, Technische Universität Kaiserslautern, D-67663 Kaiserslautern, Germany}

\author{Bert Lägel}%
\affiliation{Nano Structuring Center, Technische Universität Kaiserslautern, D-67663 Kaiserslautern, Germany}

\author{Thomas Brächer}%
\affiliation{Fachbereich Physik and Landesforschungszentrum OPTIMAS, Technische Universität Kaiserslautern, D-67663 Kaiserslautern, Germany}

\author{Carsten Dubs}%
\affiliation{INNOVENT e.V. Technologieentwicklung, D-07745 Jena, Germany}

\author{Oleksandr V. Dobrovolskiy}%
\affiliation{Faculty of Physics, University of Vienna, A-1090 Wien, Austria}

\author{Philipp Pirro}%
\email{ppirro@rhrk.uni-kl.de}
\affiliation{Fachbereich Physik and Landesforschungszentrum OPTIMAS, Technische Universität Kaiserslautern, D-67663 Kaiserslautern, Germany}

\author{Andrii V. Chumak}%
\email{andrii.chumak@univie.ac.at}
\affiliation{Faculty of Physics, University of Vienna, A-1090 Wien, Austria}

\date{\today}

\begin{abstract}
Magnonics attracts increasing attention in the view of novel low-energy computation technologies based on spin waves. Recently, spin-wave propagation in longitudinally magnetized nano-scaled spin-wave conduits was demonstrated, proving the fundamental feasibility of magnonics at the sub-\SI{100}{\nano\metre} scale. Transversely magnetized nano-conduits, which are of great interest in this regard as they offer a large group velocity and a potentially chirality-based protected transport of energy, have not yet been investigated due to their complex internal magnetic field distribution. Here, we present a study of propagating spin waves in a transversely magnetized nanoscopic yttrium iron garnet conduit of \SI{50}{\nano\metre} width. Space and time-resolved micro-focused Brillouin-light-scattering spectroscopy is employed to measure the spin-wave group velocity and decay length. A long-range spin-wave propagation is observed with a decay length of up to \SI[separate-uncertainty = true]{8.0(15)}{\micro\metre} and a large spin-wave lifetime of up to \SI[separate-uncertainty = true]{44.7(91)}{\nano\second}. The results are supported with micro-magnetic simulations, revealing a single-mode dispersion relation in contrast to the common formation of localized edge modes for microscopic systems. Furthermore, a frequency non-reciprocity for counter-propagating spin waves is observed in the simulations and the experiment, caused by the trapezoidal cross-section of the structure. The revealed long-distance spin-wave propagation on the nanoscale is particularly interesting for an application in spin-wave devices, allowing for long-distance transport of information in magnonic circuits, as well as novel low-energy device architectures.
\end{abstract}

\maketitle


Magnonics, the research field of spin-wave based data transport and information processing, aims to complement CMOS-based computation technology by replacing the charge-based binary logic with a wave-based logic \cite{Serga_2010,Khitun_2010,Kruglyak_2010,Chumak_2015,Mahmoud2_2020}. Utilizing spin waves as information carriers provides a variety of advantages such as a high energy efficiency due to the absence of ohmic losses \cite{kajiwara_2010,yu_2014,Dubs_2020} and additional degrees of freedom since frequency and phase of a spin wave are readily accessible \cite{Schneider_2008}. Moreover, spin-wave systems inherit a multitude of nonlinear mechanisms \cite{Krivosik_2010,Demidov_2009,chumak2014magnon,Sadovnikov_2016} while being scaleable to the nanoscale \cite{WangPinning_2019,Heinz_2020}, allowing for a novel device architecture \cite{Braecher_2018,Papp_2017} and simultaneously reducing the device feature size to sizes comparable or even smaller than their CMOS-based equivalent \cite{Wang_2020}. A large number of spin-wave based devices and logic elements have been realized in the recent years such as transistors \cite{Chumak_2015,Wu_2018,cramer2018magnon}, majority gates \cite{Klingler_2014,Fischer_2017,Mahmoud_2020}, directional coupler \cite{Wange1701517}, half-adder \cite{Wang_2020} or de-/multipexer \cite{Heussner_2020}, with many more theoretical concepts proposed \cite{wang2020nonlinear,wang2020inversedesign}. While being well investigated on the macro- and microscale, the study of nano-sized magnonic elements has just scratched the surface due to the limited spin-wave propagation distance on this scale and the difficulties accompanying the fabrication of nano-sized magnetic elements. Only recent progress pushed the go-to material of magnonics, yttrium iron garnet (YIG), to the nanoscale\cite{WangPinning_2019} revealing a reasonably large exponential decay length of \SI{1.8}{\micro\metre} in longitudinally magnetized YIG nano-conduits \cite{Heinz_2020}. Although a further perfection of material growth or the usage of new materials with improved characteristics might allow for an increase of the spin-wave propagation length, it is already apparent that the dominating loss channels in such systems are not intrinsic processes  \cite{gurevich1996magnetization}, but extrinsic scattering processes mediated by lattice defects, surface roughness or magnetic inhomogeneities  \cite{Heinz_2020}. Thus, novel approaches to avoid and eliminate these scattering processes are of great interest. Among them, the transport of information using topological protected (e.g. based on the Dzyaloshinskii-Moriya interaction \cite{Zhang_2013,Shindou_2013,Iacocca_2017,WangXS_2018,YamamotoK_2019}) or backscattering immune (caused by the intrinsic spin-wave mode chirality) spin-wave states is investigated. As it has been shown recently, magnetostatic surface waves (MSSW\cite{Stancil_2009}) feature such a backscattering immunity to surface defects due to the presence of energy gaps in the volume mode spectrum, rendering them insensitive to significant structural defects \cite{Mohseni_2019,MohseniM_2020}. In addition, in nanometer thick films, MSSWs are known to provide a much larger group velocity in comparison to other spin-wave modes\cite{gurevich1996magnetization,BhaskarU_2020}, which renders these waves an interesting subject of investigation regarding a long-distance data transport in magnonic circuits. However, MSSW require a transverse magnetization state leading to a strongly non-uniform internal magnetic field distribution in laterally confined nanostructures, which causes the formation of localized edge-mode states and separated volume states\cite{BayerC_2004,Gubbiotti_2004,PirroP_2014}. For nano-scaled conduits this field non-uniformity is even more pronounced and, in addition, a strong quantization is present, resulting in a complex interplay. Therefore, the mode structure in transversely magnetized nano-scaled systems is still an open question.
\newline
Here, we report on the investigation of propagating spin waves in a transversely-magnetized nano-sized YIG conduit of \SI{50}{\nano\metre} width. We show that spin waves can propagate in such a waveguide in spite of the strong non-uniformity of the internal magnetic field, observing a large spin-wave decay length up to \SI{8}{\micro\metre}. In addition, a non-reciprocity for counter-propagating spin waves is observed, which opens up the path for new device architectures. The experimental findings are supported with micro-magnetic simulations, revealing a single-mode dispersion relationship in contrast to the common formation of edge modes for microscopic systems .\newline

\begin{figure}
    \includegraphics{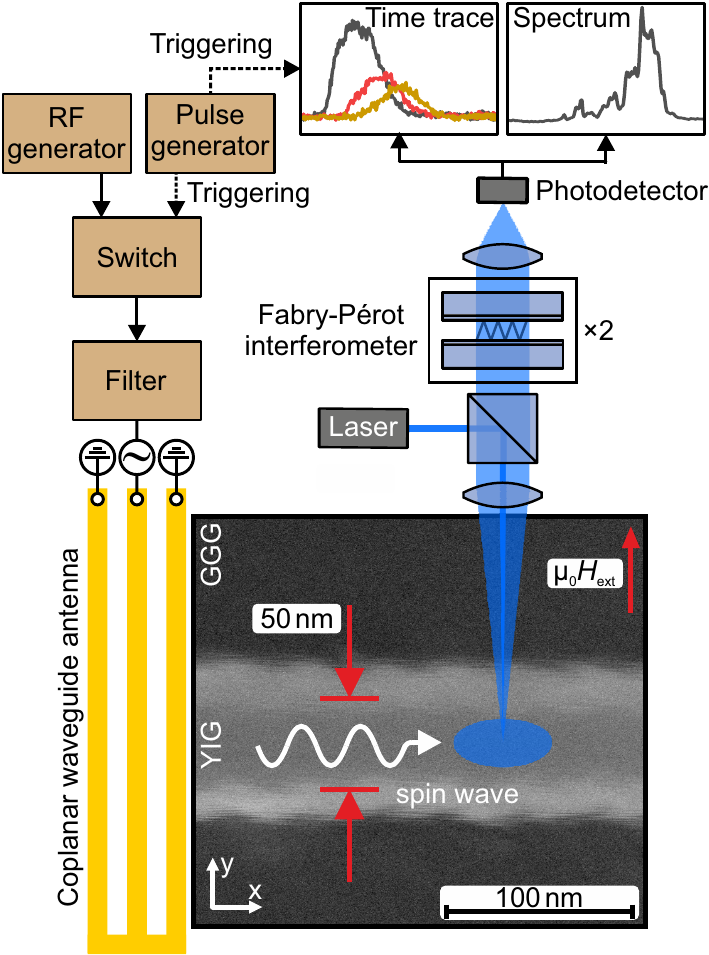}
    \caption{\label{Figure1} Experimental setup and scanning electron microscopy micrograph of the structure under investigation. A large bias magnetic field of $\mu_0 H_{\textrm{ext}}=$\SI{270}{\milli\tesla} is used to magnetize the waveguide along the short axis (\textit{y}-direction). Spin-wave packets are generated by feeding RF current pulses into a CPW antenna on top of the YIG waveguide. Frequency, spatial and time-resolved scans are performed using micro-focused BLS. The structure exhibits a trapezoidal cross-section with a top and bottom width of \SI{50}{\nano\metre} and \SI{85}{\nano\metre} respectively, caused by the fabrication process. Spin-wave wavelength and laser spot size not up to scale.}
\end{figure}

In this study, a thin lanthanum-doped \hkl(111) YIG film of \SI{44}{\nano\metre} thickness is used which is grown by liquid phase epitaxy \cite{Dubs_2017,Dubs_2020} on top of a \SI{500}{\micro\metre} thick \hkl(111) gadolinium gallium garnet substrate. A characterization of the plain film by means of vector network analyzer ferromagnetic resonance spectroscopy \cite{Kalarickal_2006,MAKSYMOV2015253} and micro-focused Brillouin-light-scattering (BLS) spectroscopy \cite{Thomas_2015} revealed the following material parameters: saturation magnetization $M_{\textrm{s}}=$\SI[separate-uncertainty = true]{140.7(28)}{\kilo\ampere\per\metre}, Gilbert damping parameter $\alpha =$\SI[separate-uncertainty = true]{1.75(8)e-4}, inhomogeneous linewidth broadening $\mu_0 \Delta H_0=$\SI[separate-uncertainty = true]{0.18(1)}{\milli\tesla} and exchange constant $A_{\textrm{ex}}=$\SI[separate-uncertainty = true]{4.22(21)}{\pico\joule\per\metre}. These parameters are within the typical range for high-quality thin YIG films \cite{Dubs_2017,Dubs_2020}. Nanoscopic waveguides were fabricated using a hard mask ion beam milling procedure \cite{Heinz_2020} resulting in conduits with a trapezoidal cross-section. The bottom of the structure is \SI{85}{\nano\metre} wide while the top width is as narrow as \SI{50}{\nano\metre}, which was determined by scanning electron microscopy, see Fig. \ref{Figure1}.
Afterwards, a gold coplanar waveguide (CPW) antenna was added on top of the waveguide with a center-to-center distance of ground and signal line of \SI{1.2}{\micro\metre}, a line width of \SI{500}{\nano\metre} and a thickness of \SI{160}{\nano\metre}. In the experiment, a large bias magnetic field of $\mu_0 H_{\textrm{ext}}=$\SI{270}{\milli\tesla} is applied in-plane along the short axis of the structure to ensure a transversely magnetized state. Radio-frequency (RF) continuous-wave (cw) currents or pulses with \SI{50}{\nano\second} length and \SI{350}{\nano\second} repetition time are generated and fed into the CPW antenna using an RF generator in combination with a fast switch and filter elements. Subsequently, propagating spin-wave packets are excited in the YIG waveguide and are detected using micro-focused BLS spectroscopy. A single-frequency laser operating at \SI{457}{\nano\metre} is used which is focused through the substrate of the sample on the structure using a compensating microscope objective (magnification $100\times$, numerical aperture $\textrm{NA} = 0.85$). In-plane spin-wave wavevectors up to $k=$\SI{24}{\radian\per\micro\metre} can be detected. The laser spot diameter is approximately \SI{300}{\nano\metre} and the effective laser power on the sample is \SI{5}{\milli\watt}.\newline
To support the findings, micro-magnetic simulations are performed using the MuMax$3$ open source framework \cite{MuMAX3}. The trapezoidal waveguide is modeled with the following dimensions: \SI{20}{\micro\metre} length, \SI{85}{\nano\metre} and \SI{50}{\nano\metre} bottom and top width respectively and \SI{44}{\nano\metre} thickness. The cell size is $10\times2.5\times5.5$\SI{}{\nano\metre\cubed}, thus introducing $8$ thickness layers with discrete varying width. The material parameters of the plain film are used in the simulations, since it has been shown that the structuring process only has a moderate influence on the structures properties \cite{Heinz_2020}. An external magnetic field of $\mu_0 H_{\textrm{ext}}=$\SI{270}{\milli\tesla} is applied in-plane along the short axis of the structure and a ground state is prepared by relaxing a random magnetization distribution. Afterwards, a driving magnetic field is applied with a spacial field distribution corresponding to the CPW antennas magnetic field to excite spin-wave packets. Time- and spacial Fourier transformations allow to extract the dispersion relationship and connected parameters.\newline

\begin{figure}
    \includegraphics{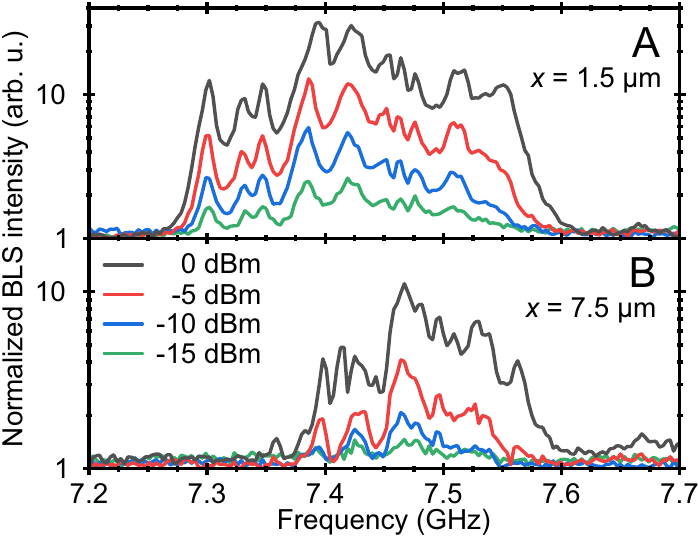}
    \caption{\label{Figure2} Spin-wave spectra in dependency of the applied microwave power. Continuous-excitation spin-wave spectra are measured at a distance of (\textbf{A}) \SI{1.5}{\micro\metre} and (\textbf{B}) \SI{7}{\micro\metre} from the CPW antenna. Low frequency states close to the ferromagnetic resonance exhibit are strongly attenuated. The spectra are normalized with respect to the thermal noise level.}
\end{figure}
\begin{figure}
    \includegraphics{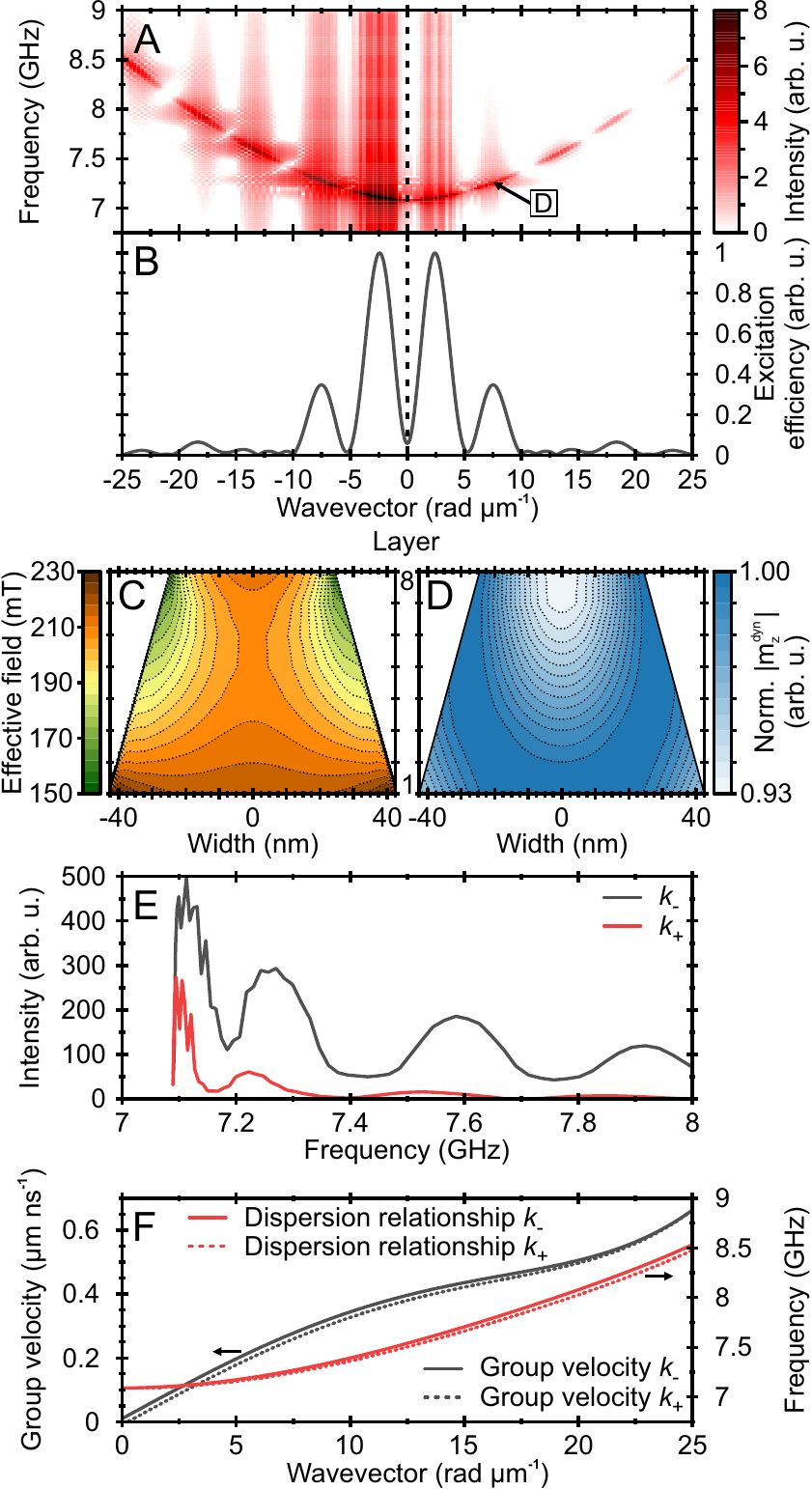}
    \caption{\label{Figure3} Micro-magnetic simulations of the investigated structure. (\textbf{A}) The dispersion relationship shows a distinct single-mode behaviour in the experimentally accessible wavevector range. (\textbf{B})  Excitation efficiency of the CPW antenna, approximated by the spacial Fourier transformation of the in-plane field distribution. The wavevector selectivity causes several gaps, visible in (\textbf{A}). (\textbf{C}) \textit{Y}-component of the internal magnetic field of the ground state. The distribution of the internal magnetic field along the \textit{y}-direction is strongly inhomogeneous. (\textbf{D}) Normalized absolute value of the dynamic out-of-plane magnetization component $\lvert\textrm{m}^{\textrm{dyn}}_{\textrm{z}}\rvert$ for a frequency of \SI{7.26}{\giga\hertz} and $k=$\SI{7.5}{\radian\per\micro\metre}. (\textbf{E}) The extracted spectral distribution shows a large intensity difference for counter-propagating spin waves. Comparison of the spectral width to Fig. \ref{Figure2} indicates that the experiment is limited to the first two excitation efficiency maxima of the CPW antenna. (\textbf{F}) A small frequency non-reciprocity for counter-propagating waves is found in the dispersion relation which is caused by the trapezoidal cross-section of the structure. The group velocity is much larger than for waves of the corresponding longitudinal magnetized state.}
\end{figure}
\begin{figure}
    \includegraphics{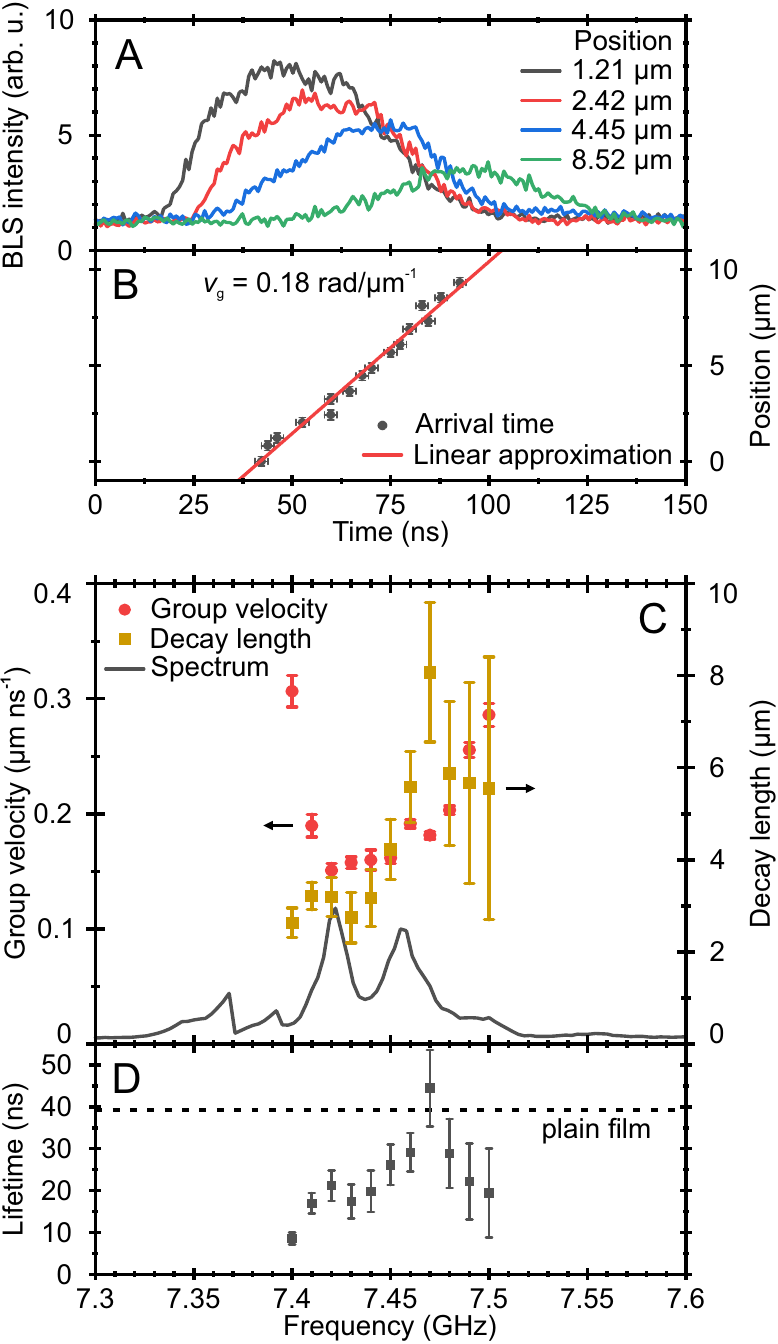}
    \caption{\label{Figure4} (\textbf{A}) Spin-wave pulses detected at different distances from the CPW antenna for a chosen excitation frequency of \SI{7.47}{\giga\hertz}. (\textbf{B}) Center-of-mass arrival time of the spin-wave pulses of (\textbf{A}). A linear approximation yields the group velocity of the wave packets. (\textbf{C}) Group velocity for various excitation frequencies extracted accordingly to the principle presented in (\textbf{A}) and (\textbf{B}). The decay length is calculated from the integrated intensity of the spin-wave pulses. The light gray curve is the associated cw excitation spectrum, slightly shifted to lower frequencies due to a magnetic field difference of \SI{0.2}{\milli\tesla}. (\textbf{D}) Experimental lifetime derived from the decay length and group velocity.  All data measured for an applied power of \SI{-10}{\dBm}.}
\end{figure}
\begin{figure}
    \includegraphics{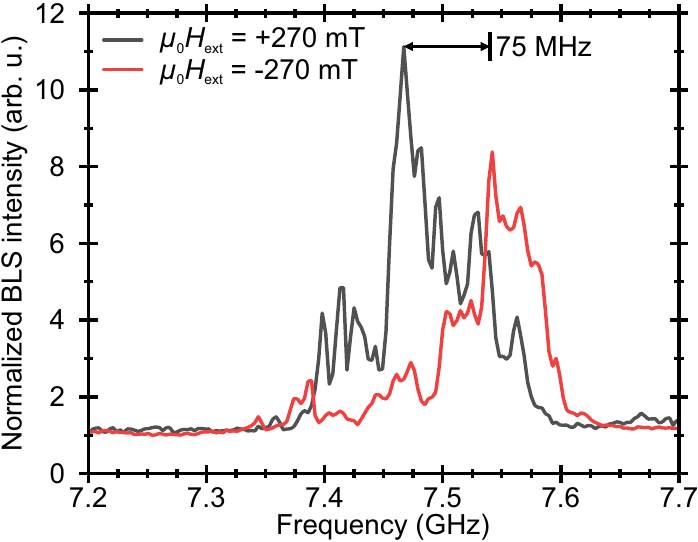}
    \caption{\label{Figure5} Frequency non-reciprocity of counter-propagating spin waves. Inverting the field polarity equals the switching of the propagation direction for the spin waves. A frequency shift of \SI{75}{\mega\hertz} is observed. The spectra are recorded for an applied power of \SI{0}{\dBm} at a distance of \SI{7}{\micro\metre} from the CPW antenna. Normalization with respect to the thermal noise level.}
\end{figure}

First, the spin-wave spectra are investigated for the case of a cw microwave excitation. The results are presented in Fig. \ref{Figure2} for different distances to the CPW antenna, \SI{1.5}{\micro\metre} (Fig. \ref{Figure2}A) and \SI{7}{\micro\metre} (Fig. \ref{Figure2}B) respectively.
The applied microwave power is varied to reveal the possible occurrence of nonlinear scattering processes which would act as additional loss channels and influence a measurement of the decay length of the system. As shown in Fig. \ref{Figure2}A, the shape of the spectrum is conserved for different applied powers, thus indicating that no strong nonlinear effects arise in the selected microwave power range. The observed spectrum has a spectral width of \SI{300}{\mega\hertz} and shows a complex behaviour, which might be attributed to microwave transmission characteristics of the used CPW antenna. Comparing the spectral distribution close to the antenna to the distribution after several micrometer of propagation (Fig. \ref{Figure2}B) reveals that the low-frequency part of the spectrum is strongly attenuated. These states are likely close to the ferromagnetic resonance of the structure and thus, as it is shown in the following, exhibit only a small group velocity.\newline
To support the findings, micro-magnetic simulations are conducted using a sinc-function pulse to realize a broadband excitation of the whole dispersion relation. In Fig. \ref{Figure3}A the resulting excited spectrum is shown for the experimentally accessible wavevector range.
A clear single mode state is observed following a monotonous function. However, the absolute frequency is shifted by approximately \SI{300}{\mega\hertz} with respect to the experimental results of Fig. \ref{Figure2}, which indicates that the magnetic parameters are slightly alternated by the structuring process \cite{Heinz_2020}. In addition, a small variation of the magnetic width of the structure can also significantly impact the spin-wave frequency on the present scale. The various small gaps in the dispersion relationship are caused by the wavevector selective excitation efficiency of the CPW antenna, as shown in Fig. \ref{Figure3}B. Here, the efficiency is approximated by the spacial Fourier transformation of the in-plane field distribution, for a detailed discussion see \cite{Demidov2_2009}. The simulated internal field distribution of the ground state is shown in Fig. \ref{Figure3}C. As expected, large demagnetizing fields arise leading to a strongly non-uniform internal field distribution along the external magnetic field direction. Similar to micron-sized conduits distinct regions of a significantly reduced effective magnetic field are formed at the edges of the structure. However, since the structure is too small to allow for a homogeneous field region in the center, a single mode behaviour is observed in contrast to the typically observed appearance of localized edge mode states. This is validated by the mode profile (normalized absolute value of the dynamic out-of-plane magnetization component $\lvert\textrm{m}^{\textrm{dyn}}_{\textrm{z}}\rvert$) shown in Fig. \ref{Figure3}D for a frequency of \SI{7.26}{\giga\hertz} and $k=$\SI{7.5}{\radian\per\micro\metre}, which extents fully into the edge regime of the structure.\newline
Extracting the respective spectral distribution, see Fig. \ref{Figure3}E, shows an intensity difference for counter-propagating waves ($k_{-}$ and $k_{+}$), which is caused by a non-reciprocal excitation efficiency of MSSW when using a microwave antenna\cite{Schneider_2008,Demidov2_2009}. Moreover, comparing the spectra to Fig. \ref{Figure2} indicates that the experiment is limited to the first two excitation efficiency maxima of the CPW antenna. Excitation maxima of higher order are likely not observed due to the limited sensitivity of the used BLS setup, especially due to the small amount of probed material and the background of thermal spin waves leading to an increased noise level in the experiment.\newline
In Fig. \ref{Figure3}F the dispersion relationship, derived from a 6\textsuperscript{th}-order polynomial approximation of Fig. \ref{Figure3}A, and the a group velocity, extracted as the derivative of the dispersion relationship, are shown. Similar to microscopic systems, the transversely magnetized state offers a high and much larger group velocity than the waves of the corresponding longitudinal magnetized state possess \cite{Heinz_2020}.\newline 
In the following, the group velocity and the decay length are determined experimentally by a direct measurement of propagating spin-wave packets to further characterize the system and investigate whether a long range spin-wave propagation can be realized in such nano-scaled systems. Thus, a pulsed excitation with \SI{50}{\nano\second} pulse length and \SI{350}{\nano\second} repetition time is used, choosing an applied microwave power of \SI{-10}{\dBm} to ensure an operation within the linear regime. In Fig. \ref{Figure4}A time-resolved BLS measurements of the excited spin-wave packets are shown for different positions along the conduit.
The respective center-of-mass arrival time of the pulse is determined by subtracting the thermal noise and calculating the weighted average of the packet. A linear approximation of the packet arrival time, as shown in Fig. \ref{Figure4}B, yields the group velocity $v_{\textrm{g}}$. Additionally, the decay length can be extracted from the integrated pulse intensity of each position by approximating the decay as follows:
\begin{equation}
     I=I_{1}\exp(-2x/\lambda_{\textrm{D}})+I_{0}.
\end{equation}
Here, $I_{1}$ denotes the initial intensity, $I_{0}$  the offset intensity, $x$ the position, and $\lambda_{\textrm{D}}$ the decay length. Extracted accordingly to this principle, the resulting group velocities and decay lengths are presented in Fig. \ref{Figure4}C. The velocity lies within the expected range predicted by the simulations when compared for wavevectors up to \SI{10}{\radian\per\micro\metre}, see Fig. \ref{Figure3}F, and follows the rising trend to higher frequencies. However, an unexpected large velocity is observed for the smallest investigated frequency not covered by the prediction of the simulations. In contrast, the decay length follows a steep increase up to a maximum of \SI[separate-uncertainty = true]{8.0(15)}{\micro\metre}, slightly dropping off for higher frequencies. For comparison, the corresponding cw excitation spectrum is displayed in light gray, which is, however, slightly shifted to smaller frequencies due to a small magnetic field difference of \SI{0.2}{\milli\tesla}. Taking this shift into account, the second spectral peak matches the frequency of the maximum decay length. We would like to point out, that the observed decay length of \SI{8.0}{\micro\metre} is significantly larger in comparison to reported values in the longitudinal magnetization configuration of \SI{1.8}{\micro\metre}\cite{Heinz_2020}. This potentially enables complex nano-sized integrated spin-wave circuits consisting of multiple elements without any means of intermediate amplification, severely lowering the energy consumption of such circuits.\newline
In the following, the lifetime $\tau$ of the propagating waves, shown in Fig. \ref{Figure4}D, is derived from the experimental results using the expression
\begin{equation}
     \tau=\lambda_{\textrm{D}}/v_{\textrm{g}}.
\end{equation}
Here, a rather large lifetime of up to \SI[separate-uncertainty = true]{44.7(91)}{\nano\second} is found. Calculating the lifetime of the ferromagnetic resonance of the plain film as a comparison\cite{Stancil_2009}, considering the decreased internal magnetic field of the transverse magnetization state, results in \SI{112.5}{\nano\second} for only Gilbert type losses and \SI{39.2}{\nano\second} taking the full linewidth into account. Here, two effects likely take place affecting the considerations: On the one hand, the structuring procedure influences the material parameters\cite{Heinz_2020} and potentially reduces the lifetime compared to the plain film. On the other hand, the non-local inhomogeneous linewidth broadening extracted from the plain film overestimates the effective inhomogeneity on the length scale of the nanostructure\cite{Hahn_2014}. Thus, a comparison to the full linewidth of the plain film can yield a lifetime smaller than the lifetime of the nanostructure. Nonetheless the measured lifetime exceeds other commonly used materials such as permalloy\cite{Yamanoi_2013} by far, even when comparing to microscopic or macroscopic systems.\newline
Finally, we would like to discuss a peculiarity of the investigated system, observed in Fig. \ref{Figure3} F. The simulated dispersion relationship exhibits a small frequency non-reciprocity, which is caused by the trapezoidal cross-section of the structure and the associated internal field distribution (Fig. \ref{Figure3}C). This introduces an additional spacial symmetry break, similar to the case of magnetic bilayers \cite{GrassiM_2020}. In Fig. \ref{Figure5} two measured spectra for normal and inverted field polarity are shown, which equals a switch of the dispersion branch from $k_{-/+}$ to $k_{+/-}$.
Indeed, a distinct non-reciprocity with a frequency shift of \SI{75}{\mega\hertz} is found, which is substantially larger than the predicted shift of \SI{20}{\mega\hertz}\textendash\SI{30}{\mega\hertz} for spin waves in the range of \SI{7}{\radian\per\micro\metre}\textendash\SI{10}{\radian\per\micro\metre}, see Fig. \ref{Figure3}F. It should be noted that, the observed frequency shift is likely influenced by the non-reciprocal excitation efficiency for both configurations, leading to different spin-wave densities, and thus to a different nonlinear frequency downshift potentially increasing the observed frequency gap. Nonetheless, such a pronounced non-reciprocity is of particular interest for an application in spin-wave devices since it allows for a novel device architecture, e.g. allows for the construction of a nano-sized spin-wave diode.
 \newline


To conclude, we presented a study of propagating spin-wave packets in a transversely magnetized nano-scaled YIG conduit of \SI{50}{\nano\metre} width. Micro-magnetic simulations are performed to support the experimental findings, revealing a single-mode dispersion relationship in contrast to the common formation of localized edge modes for microscopic systems. It is shown that the  observed mode is not localized within the central part of the structure. A large spin-wave group velocity is measured and a long-range spin-wave propagation is observed with a decay length of up to \SI[separate-uncertainty = true]{8.0(15)}{\micro\metre}, which is multiple times larger than reported values for the corresponding longitudinal magnetized state\cite{Heinz_2020}. In addition, a large spin-wave lifetime of up to \SI[separate-uncertainty = true]{44.7(91)}{\nano\second} is found. Furthermore, a frequency non-reciprocity for counter-propagating spin waves is observed and experimentally verified, which is caused by the trapezoidal cross-section of the structure and the associated internal field distribution. This non-reciprocity and the revealed long-distance spin-wave propagation on the nanoscale are particularly interesting for an application in spin-wave devices, allowing for long-distance transport of information in magnonic circuits, as well as novel low-energy device architectures potentially opening up the path to multi-element circuits without intermediate amplification.\newline\newline

This research has been funded by the European Research Council project ERC Starting Grant 678309 MagnonCircuits, by the Deutsche Forschungsgemeinschaft (DFG, German Research Foundation) - 271741898, by the Collaborative Research Center SFB/TRR 173-268565370 (Project B01), and by the Austrian Science Fund (FWF) through the project I 4696-N. B.H. acknowledges support from the Graduate School Material Science in Mainz (MAINZ). The authors thank Burkard Hillebrands for support and valuable discussions.

\nocite{*}

\bibliography{ManuscriptDEWaves}

\end{document}